# Design, Testing and Numerical Modelling of a Low-Speed Wind Tunnel Gust Generator

Marinos Manolesos[1], Christos Ampatis[1], Dimitris Gkiolas[1], Konstantinos Rekoumis[2] and George Papadakis[2]

[1]*School of Mechanical Engineering, National Technical University of Athens, Zografos, Athens, 15780, Greece*
[2]*School of Naval Architecture and Marine Engineering, National Technical University of Athens, Zografos, Athens, 15780, Greece*

**ABSTRACT**

Accurate reproduction of deterministic gusts in wind tunnels is essential for studying unsteady aerodynamics and aeroelastic response in aircraft, uninhabited aerial vehicles, and wind turbines. This work presents the design, experimental characterisation, and numerical modelling of a low-speed gust generator based on oscillating vanes, capable of producing high-amplitude gusts in strongly unsteady flow regimes. Cross-flow hot-wire measurements are combined with time-accurate computational fluid dynamics simulations to analyse gust formation and propagation. Classical '1-cos' gusts are shown to exhibit pronounced negative velocity peaks associated with start–stop vortex shedding. A modified vane motion protocol is proposed that significantly reduces the negative peak factor while preserving a substantial gust ratio over a wide range of reduced frequencies. Flow-field analysis reveals that secondary variations in gust angle arise from nonlinear interactions between vortices shed by adjacent vanes and are influenced by wind-tunnel confinement. The results provide physical insight into the limitations of oscillating-vane gust generators and guidance for the design of high-fidelity gust-generation systems.

## 1  Introduction

Gusts of moderate and large magnitude can lead to flow separation, transient stall, and other unsteady phenomena when they interact with lifting surfaces on air vehicles or large wind turbines. In contemporary aerospace and wind-energy applications, such gust encounters are unavoidable, and the classical potential-flow-based methods traditionally used to predict unsteady aerodynamic response are no longer adequate for capturing the associated non-linearities. A deeper understanding of the physical mechanisms governing gust related interactions is therefore essential both for accurate load prediction and for the development of effective gust-mitigation and load-alleviation strategies. Achieving this requires controlled, accurate and repeatable experimental data. The motivation for the present work is thus to provide a high-fidelity, mechanically robust gust-generation capability that enables systematic investigation of unsteady aerodynamic response under controlled conditions, supporting both fundamental research and the development of next-generation engineering analysis methods.

The remainder of this paper is organised as follows. Section 2 reviews deterministic gust models relevant to aircraft, UAVs, and wind turbines, and summarises existing approaches to gust generation in low-speed wind tunnels, with emphasis on oscillating-vane concepts. Section 3 describes the design, manufacturing, and installation of the new gust generator, together with the experimental setup, measurement methodology, and the gust profiles examined. Section 4 presents the experimental results and discusses the influence of key parameters such as free-stream velocity, vane amplitude, excitation frequency, and motion protocol. Numerical simulations are then used to validate the measurements and to provide further insight into gust propagation and vortex dynamics. Finally, Section 5 summarises the main findings and outlines the implications of the present work for future experimental and numerical studies of unsteady aerodynamics.

# 2 Background

In this section a short review of gusts relevant to Aircraft, Uninhabited Air Vehicles (UAVs) and Wind turbines is presented followed by a review of mechanisms used to generate gusts in low-speed wind tunnel facilities.

## 2.1 Gusts on Aircraft

In the past, many deterministic gust shapes have been used to assess wind loading on aircraft. Many of these classical problems are posed by employing certain wave functions, such as Küssner's problem, see Figure 1, top. Another common example of a deterministic gust is the '1-cos' gust, sketched in Figure 1, middle, which is part of the Federal Aviation Administration (FAA) regulations [1].

The '1-cos' type gust is used in aircraft certification for dynamic load requirements within the gust envelope of the aircraft. The European Aviation Safety Agency's Certification Specifications address the gust profile in the same way for the following aircraft categories: Large Civil Aircraft [2], Normal, Utility, Aerobatic, and Commuter Aircraft [3] and Very Light Aircraft [4].

Hoblit [5] states that there is a wide range of frequencies at which an aircraft might encounter an atmospheric gust. Thus, in order to investigate these kinds of applications, it is necessary to make sure that the gust generation mechanism is capable of functioning over a wide frequency range during the pre-design phase. Achieving a wide range of frequencies is also significant because, in aeroelasticity experiments, the response of the system depends heavily on the region where the excitation frequency approaches the natural frequencies of the structure.

## 2.2 Gusts on Uninhabited Aerial Vehicles

Regulations governing the certification of UAVs in Europe remain lacking, in contrast to those of manned aircraft. In this sense, there are no standardised specifications for the gust envelope of these structures.

In some cases, the amplitude of the gust disturbance in UAVs is of the same order of magnitude as the flight speed. Jones & Cetiner [6] reviewed a set of experiments and simulations involving large-amplitude transverse gust encounters (with gust ratios ranging from 0.2 to 2.6), revealing the formation of a strong leading-edge vortex, which is the primary cause of an observed lift peak, and the subsequent flow separation. Gust ratio (GR) is defined as the ratio of the vertical velocity, $v$, to the free stream velocity, $U_\infty$, see equation (1).

$$GR = \frac{v}{U_\infty} \qquad (1)$$

## 2.3 Gusts on Wind Turbines

Similar to aircraft specification regulations, the '1-cos' velocity distribution is also utilized in the international standard for wind turbines [7], specifically referred to as Extreme Coherent Gust with direction change, with a magnitude of 15 m/s and a rise duration of T = 10 seconds.

Furthermore, the international standard specifies the Extreme Operating Gust disturbance. The latter, is notoriously difficult for control systems to handle [8], has a 'Mexican hat' profile (see Figure 1, c) and can be expressed mathematically as follows:

$$u'(t) = \begin{cases} -0.37 A \sin\left(\frac{3\pi t}{\tau}\right)\left[1 - \cos\left(\frac{2\pi t}{\tau}\right)\right], & for\ 0 \leq t \leq \tau \\ 0, & otherwise \end{cases} \qquad (2)$$

where the time scale, τ, is taken as 10.5 seconds.

Although deterministic gust models may deviate from real atmospheric conditions, due to the assumption of uniform inflow [9], the goal of the standard is to confirm that any new wind turbine design can withstand even the harshest loading conditions over the course of its lifetime.

Under normal operating conditions a portion of the wind turbine blade experiences some degree of stall. The inflow of the rotating blade is inherently unsteady suggesting that unstable aerodynamics, particularly dynamic stall, are crucial for wind turbine applications and their control strategies [10]. The gust in some cases results in flow separation and shedding of vorticity in the wake of the wind turbine blade, similar to a dynamic stall event of a pitching airfoil [9]. Despite the fact that the foregoing has compelled researchers to develop methods to alleviate gust-induced loads, there is a significant lack of experimental works that concern low Mach number applications, similar to those found in wind turbines.

Exception to the above constitutes the recent wind tunnel work carried out at the Technical University Braunschweig [11]. The researchers suggest a characterization of the dynamic lift hysteresis of an airfoil with a pitching trailing edge flap at a chord Reynolds number Re = $1.8 \times 10^6$ by experimental measurements. The same team also conducted measurements of the lift response [12] in order to ultimately suggest a closed-loop control strategy for vortex-gust disturbances.

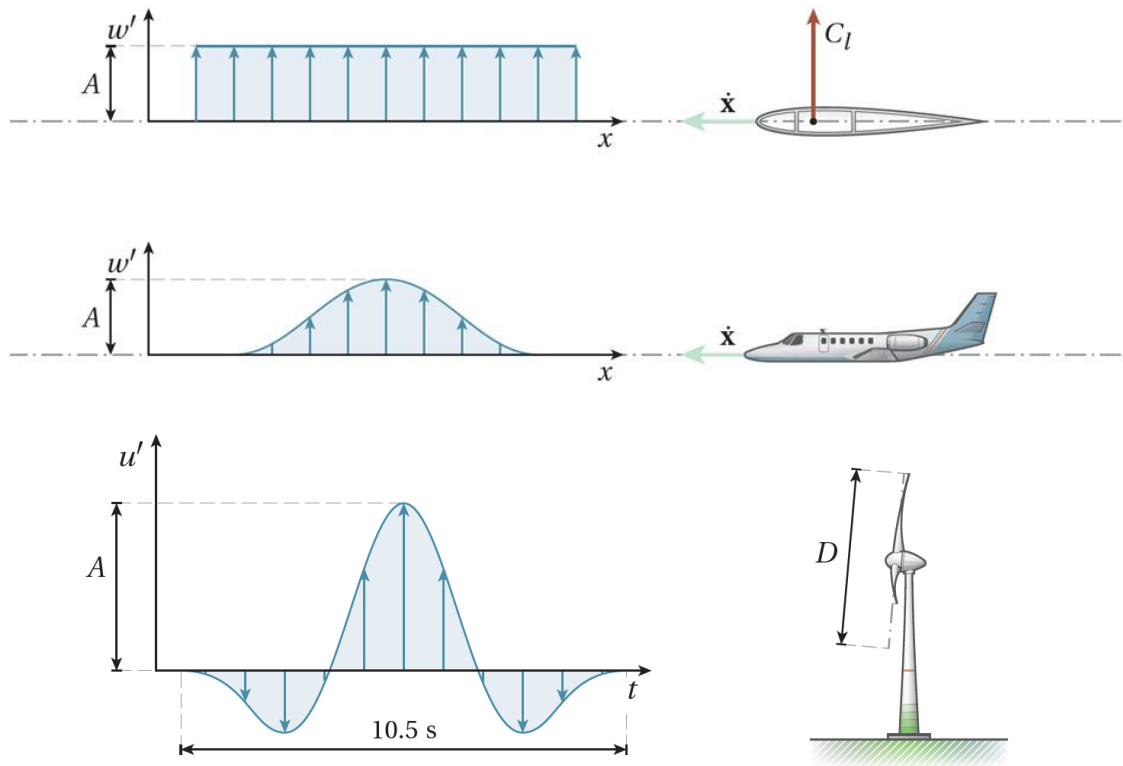

Figure 1. Examples of deterministic gusts models in engineering. Top: Kuessner's problem; Middle: '1-cos' gust; Bottom: Extreme operating Gust. Adapted from [9]

## 2.4 Gusts Generation in Wind Tunnels

Gust generation in wind tunnels has been approached using a variety of passive and active concepts. Early designs relied on fixed devices at the inlet of the test section to alter the mean flow and generate a prescribed turbulent field, but such passive methods are limited to statistically steady turbulence and cannot reproduce discrete, high-amplitude gust events. Active concepts were therefore introduced, based on the controlled motion of aerodynamic surfaces.

Examples include rotating slotted cylinders placed downstream of fixed vanes [13,14], active Makita-type grids using independently actuated flaps [15–19], circulation-control airfoils [20–22], shutter-type plates upstream of the test section [23,24], deformable "wavy" walls at the inlet [25], and multiple-fan active-control installations [26]. Each of these configurations offers a different balance between complexity, achievable gust amplitude, spectral content, and flow uniformity. Among them, however, systems based on oscillating vanes mounted at the test-section inlet have emerged as one of the most widely adopted methods in university and research-facility applications for their simplicity, effectiveness and flow quality [27–29].

One of the earliest oscillating-vane Gust Generators (GGs) was designed and built at NASA Ames Research Center in 1969 [30]. The installation comprised multiple NACA 0015 vanes arranged in several configurations with a vertical spacing of one chord length. A 125 hp, six-pole induction motor drove the vanes in periodic motion with an amplitude of 16° and frequencies between 1 and 15 Hz. Experiments demonstrated that lateral gust components up to 45% of the free-stream velocity (GR=0.45) could be generated, although accurate control of the gust angle remained challenging. This work established the feasibility of producing strong, repeatable gusts in a low-speed wind tunnel using oscillating vanes.

A subsequent implementation at the Virginia Tech Stability Wind Tunnel introduced a computer-controlled GG with ten vertical vanes of modified NACA 0018 profile spanning the contraction height [31]. Independent stepper motors actuated each vane, enabling both periodic and discrete disturbances to be prescribed. The system achieved GRs of approximately 0.35 at frequencies up to 25 Hz, with good repeatability and consistent waveform shape. This configuration demonstrated the benefits of individually controlled vanes for tailoring the spatial and temporal structure of the gust.

[32] investigated a single oscillating vane installed in the transonic wind tunnel at Göttingen. A hydraulically actuated NACA 0010 airfoil was driven to large amplitudes and frequencies, reaching up to 20° and 60 Hz. Experimental measurements were used to validate high-fidelity Computational Fluid Dynamics (CFD) simulations based on the compressible, time-accurate RANS equations [32]. The study highlighted the importance of mesh refinement around the leading and trailing edges and in the wake region to capture the complex unsteady flow structures generated by the vane. However, later work by Lancelot et al. [33] pointed out that single-vane configurations introduce strong shear regions in the wake, which adversely affect flow uniformity around the test model. To address this issue, an even number of vanes is commonly used, to ensure flow uniformity and sufficiently high gust strength [14,27–29,33–36].

A comprehensive design study for a GG to be installed in the large wind tunnel of Politecnico di Milano is described in [27]. Two-dimensional unsteady CFD simulations explored configurations with 2, 6, and 8 vanes, chord lengths between 200 and 400 mm, and deflections ranging from 6° to 12°. The final design comprised six NACA 0012 vanes of 400 mm chord, spanning the full test-section height. This study emphasised the role of vane number, chord length, and deflection amplitude on the generated gust field and blockage.

A similar combined numerical–experimental approach was followed by [14] for the wind tunnel at the University of Bristol. There, a pair of horizontal NACA 0015 vanes were selected after a parametric Reynolds Averaged Navier Stokes (RANS) CFD study. The experimental results revealed that at vane frequencies above about 8 Hz the nominally sinusoidal gusts became increasingly distorted and approached a more square-wave form, underlining the importance of structural stiffness and actuator bandwidth at higher frequencies.

The two-vane concept has since been adopted in several additional facilities. At the University of Maine Crosby Hall wind tunnel, a pair of NACA 0018 vanes with 150 mm chord were separated by

230 mm and driven synchronously by a stepper motor [28]. CFD simulations using a uniform inlet flow and the k–ω SST turbulence model indicated a velocity-deficit region and elevated turbulence levels in the vane wake, which gradually diminished over a distance of roughly ten chords downstream.

Building on this accumulated experience, a two-vane GG was designed and installed in the Swansea University wind tunnel, incorporating several refinements from previous systems [34]. The vanes were NACA 0015 sections with 200 mm chord, constructed from eight 3D-printed segments mounted on a steel spar. Independent servomotors with low-backlash gearboxes actuated each vane, substantially reducing inertial and torque requirements. Experiments demonstrated both discrete and continuous gusts with frequencies up to 14 Hz and amplitudes between 5° and 20°. In this work, the authors further demonstrated that the generation of nominal discreet gusts (e.g. 1-cos) is not trivial and requires specific vane motion protocols.

Table 1 summarizes the technical characteristics of GG setups involving oscillating vanes that have been installed in low-speed wind tunnels worldwide. First, the direction normal and parallel to the vane span (A and B, respectively) is given, along with the maximum freestream velocity of the wind tunnel facility. Next, the design parameters of the GG system are stated, including the vane profile. The defined ratios provide a dimensionless view of each configuration, enabling the comparison between setups installed in facilities of different size.

## 2.5 The Present Study

The literature suggests a clear evolution in oscillating-vane GGs. From early, mechanically driven multi-vane systems towards computer-controlled, multi-axis servomotor installations with carefully optimised vane geometry, spacing, and actuation protocols. The GG described in this work is directly informed by these developments, targeting high gust amplitudes, broad frequency range, good flow uniformity, and mechanical robustness suitable for detailed unsteady aerodynamic investigations. In addition, we propose a modified motion protocol to reduce the negative velocity peaks that deviate from the nominal '1-cos' gust. The proposed motion is compared to the nominal motion and analysed using CFD. The study reveals how nonlinear vortex–vortex interactions between adjacent vanes fundamentally limit the fidelity of deterministic gusts in confined domains.

The novelty of the present study lies in (i) the combined experimental–numerical quantification of negative peak mitigation, and (ii) the identification of vortex-interaction mechanisms that fundamentally limit gust fidelity in oscillating-vane systems.

This paper continues with an overview of the gust generator design and manufacturing, followed by a description of the experimental setup. Experimental results are then presented, followed by the numerical investigation, and the study concludes with a summary of the main findings.

Table 1. Summary of vane-type low-speed gust generator design parameters from the literature

| Institution | A (m) (normal to vane span) | B (m) (parallel to vane span) | $U_{max}$ (m/s) | N (# of vanes) | Vane profile (NACA) | c (m) | B/c | c/A | N·c/A | Total Blockage |
|---|---|---|---|---|---|---|---|---|---|---|
| *NASA Ames (1968)* [30] | 3.05 | 2.13 | 76 | 6 | 0015 | 0.146 | 14.6 | 0.05 | 0.29 | 4.3% |
| *Virginia Tech (2004)* [31] | 2.15 | 2.15 | 80 | 10 | 0018 | 0.138 | 15.6 | 0.06 | 0.64 | 11.6% |
| *Cranfield Univ. (2015)* [35] | 1.52 | 1.14 | 32 | 6 | 0015 | 0.114 | 10.0 | 0.08 | 0.45 | 6.8% |
| *POLIMI (2016)* [27] | 4.00 | 3.84 | 55 | 6 | 0012 | 0.400 | 9.50 | 0.10 | 0.60 | 7.2% |
| *TU Delft (2017)* [33] | 2.85 | 2.85 | 35 | 2 | 0014 | 0.300 | 9.50 | 0.11 | 0.21 | 2.9% |
| *Univ. of Bristol (2017)* [14] | 1.52 | 2.14 | 60 | 2 | 0015 | 0.300 | 7.13 | 0.20 | 0.39 | 5.9% |
| *Univ. of Maine (2021)* [28] | 0.75 | 0.75 | 24 | 2 | 0018 | 0.150 | 5.00 | 0.20 | 0.40 | 7.2% |
| *METU (2022)* [29] | 0.34 | 0.34 | 25 | 2 | 0015 | 0.080 | 4.25 | 0.24 | 0.47 | 7.1% |
| *Swansea Univ. (2022)* [34] | 1.00 | 1.5 | 50 | 2 | 0015 | 0.200 | 7.50 | 0.20 | 0.40 | 6.0% |
| *Univ. of Arizona (2024)* [36] | 0.91 | 1.22 | 80 | 2 | 0015 | 0.174 | 7.00 | 0.19 | 0.38 | 5.7% |
| *NTUA (2025, present study)* | 1.80 | 1.40 | 60 | 4 | 0015 | 0.200 | 7.00 | 0.11 | 0.44 | 6.7% |

# 3 Methods

## 3.1 Experimental set up

### 3.1.1 Wind Tunnel Test Section

The National Technical University wind tunnel is a closed-loop, closed test section facility. The small wind tunnel test section has a width of 1.8 m, a height of 1.4 m and a total length of 3.2 m. The maximum velocity can reach 60 m/s, while the turbulence intensity is 0.2%. The new GG is designed to be combined with the existing aeroelastic set up, where the wing models are mounted vertically [37], so the vanes are also mounted vertically at the entry of the test section.

### 3.1.2 Gust Generator Design

Based on the existing literature and the considered applications (Aircraft, UAV, Wind Turbines) the specifications for the GG were defined as given in Table 2. The GG consists of four vertically oriented NACA 0015 vanes with a chord of 0.20 m and a span of 1.4 m, i.e. spanning the tunnel height. The system is designed to operate within an envelope that accommodates oscillation frequencies up to 20 Hz and vane deflections up to ±20°, enabling the reproduction of a broad range of prescribed gust profiles.

The GG specifications require a maximum GR of at least 0.3 with as low Negative Peak Factor (NPF) as possible. The latter is a measure of the agreement of the achieved gust profile to the nominal '1-cos' gust, as suggested in [34]. NPF is calculated according to equation (3):

$$NPF = \left|\frac{\theta_{min}}{\theta_{max}}\right| \qquad (3)$$

where $\theta_{min}$ and $\theta_{max}$ are respectively the minimum and maximum values of the gust angle during the produced gust. The objective is therefore the decrease of the $NPF$ value, which is accomplished by reducing the negative peaks while keeping the maximum achieved GR high.

Each vane is actuated by an independent servomotor equipped with a low-backlash, high-stiffness planetary gearbox with a 10:1 reduction ratio, providing the torque and positional accuracy required for high-fidelity motion control. The motors deliver a maximum torque of 8.5 Nm and a continuous rating of 2.8 Nm, ensuring robust operation across the full actuation range.

Further to other technical requirements, the installation procedure was developed for consistent repeatability, allowing the vane positions to be reinstalled with high angular accuracy after each assembly. At the same time, given the generic nature and use of the wind tunnel test section, the assembly is mounted on a modular support structure that allows installation or removal within approximately two hours, facilitating efficient integration into the wind-tunnel test section. A rendered view of the final design is given in Figure 2.

Table 2. Specifications for the NTUA wind tunnel GG

| Category | Parameter | Value |
|---|---|---|
| Geometry | Number of Vanes | 4 |
| | Vane chord | 0.2 m |
| | Vane Profile | NACA 0015 |
| | Vane Span | 1.4 m |
| | Material | 3D printed (ABS) |
| Operation | Max Frequency | 20 Hz |
| | Max Angle range | 20° |
| Actuation | Geared Servomotor | One motor per Vane |
| | Planetary Gearbox | *Backlash <0.05°/ Rigidity 0.1°@30Nm* |
| | Gearbox Ratio | 10:1 |
| | Motor Power | 750W |
| | Motor Torque | 8.5 Nm (max) / 2.8 Nm (continuous) |
| Performance | Maximum Gust Ratio | $\geq$ 0.25 at max |
| | Negative Peak Factor | As low as possible |
| Installation | | *Install / Uninstall within 2 hours* |

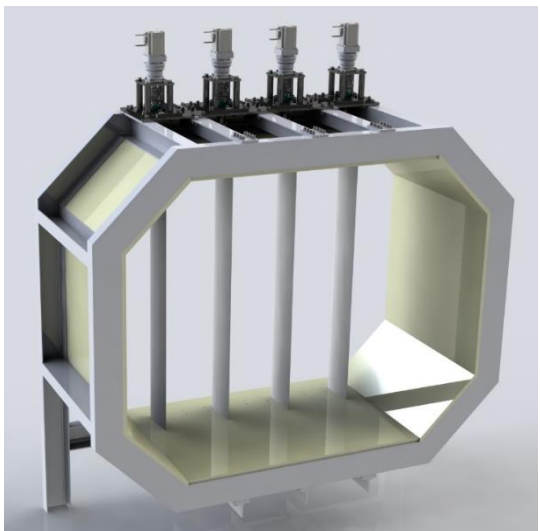

Figure 2. Render of the final GG design

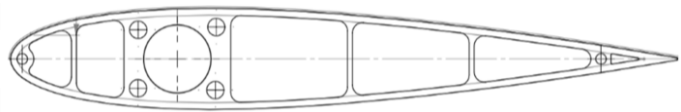

Figure 3. Drawing of the vane profile

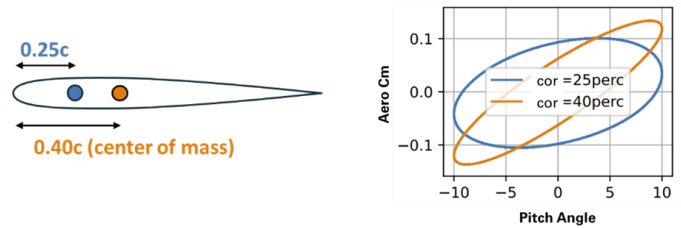

Figure 4. Aerodynamic moment coefficient of the gust generator vane for different points of rotation. CFD results for $f = 20\,Hz$ and $A = \pm 10°$.

Compared to similar designs in the literature, the NTUA wind tunnel GG has a 'relevant total chord' (the product of number of vanes times the chord length divided by the tunnel width) on the average of the relevant literature for a given blockage, see Figure 5-left. At the same time, the design achieves a lower chord to tunnel width ratio for a given vane aspect ratio, see Figure 5-right, by employing a larger number of vanes. This leads to a smaller vane inertia and enables the GG to achieve higher accelerations.

To minimise vane inertia, the vanes were designed to achieve minimum mass while maintaining maximum rigidity. The final vane profile is shown in Figure 3. The quarter-chord point (c/4) was selected as the axis of rotation. Although this choice increases the moment of inertia, it significantly reduces variation in the aerodynamic moment, see Figure 4, resulting in a more controllable disturbance. The analysis showed that this design choice led to only a marginal increase of approximately 1% in the required motor torque.

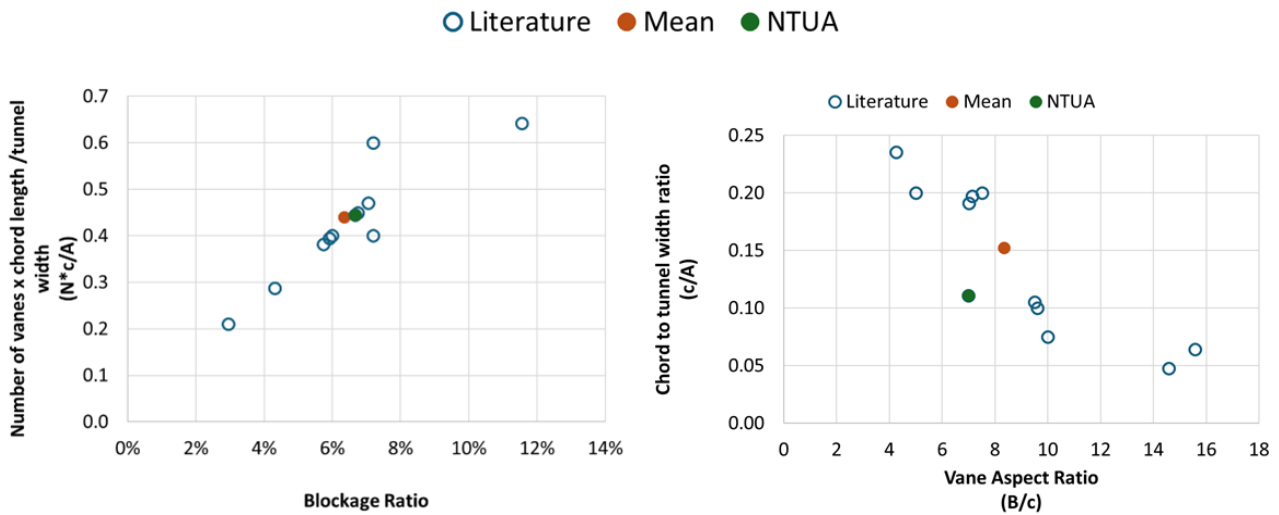

Figure 5. Gust generation design characteristics. Total chord length vs blockage ratio (left) and chord to width ratio versus aspect ratio (right). Values form the literature and from the present design. The literature mean value is shown in green.

### 3.1.3 Gust Generator Manufacturing

The vanes were manufactured using Fused Deposition Modelling 3D printing in ABS to minimise weight, and hence inertia, while preserving structural rigidity. They were printed in 7 different parts of equal chord but varying spanwise length (5 longer middle pieces, 2 tip pieces). The vane pieces were bonded on a single hardened steel shaft of 0.2 m diameter. To improve surface finish, the vanes were polished and painted post assembly, see Figure 6. The total vane mass is approximately 6 kg.

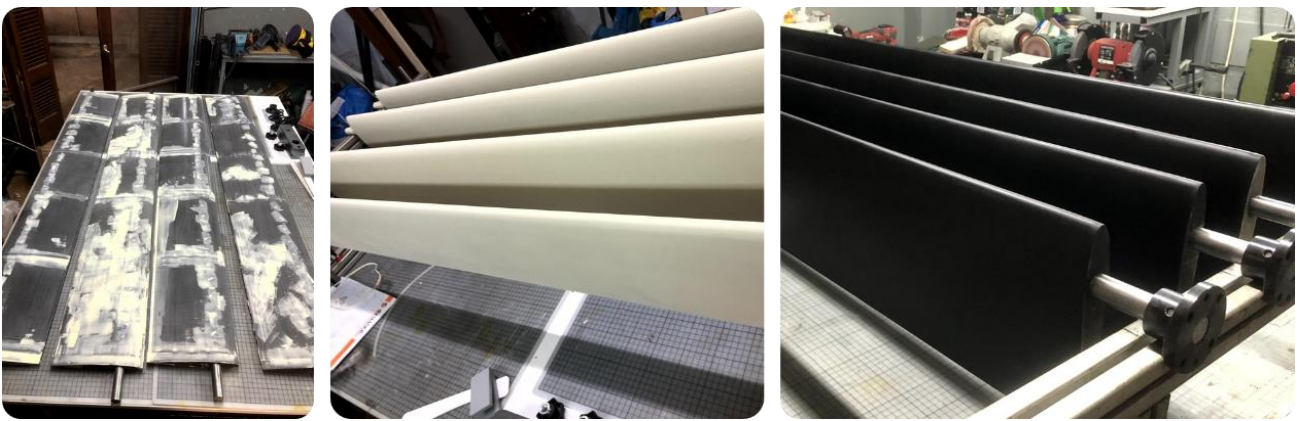

Figure 6. Manufacturing the vanes. From sanding the assembled vanes (left) to the final painted result (right).

Having a solid shaft instead of hollow, had a minimal effect on the total vane inertia, but significantly simplified manufacturability and improved rigidity. Analytical estimates and finite-element calculations, not presented here in the interest of brevity, confirmed that torsional deflection remains negligible across the full operational frequency range. The shaft is supported by a pair of back-to-back angular-contact bearings housed within a precision-machined support block. This arrangement provides high radial and axial stiffness, ensuring that off-axis loads induced by aerodynamic forcing do not degrade transmission fidelity.

Transmission of torque from the motor to the vane shaft is achieved using zero-backlash flexible couplings. These couplings provide angular compliance for small misalignments during installation, while maintaining high torsional stiffness. The absence of measurable backlash is essential for reproducing high-frequency gust motions, as even minor mechanical play would introduce phase error, amplitude attenuation, and trajectory discontinuities. All shaft-to-coupling connections use clamping interfaces to ensure repeatable installation without the need for re-machining or bearing preload adjustment.

Inertia matching between the motor, $J_M$, and the vane assembly, $J_L$, was a key requirement. The combined inertia of the vane, shaft, bearings, and couplings was evaluated through CAD-based computation and validated against motor-controller measurements during commissioning. The resulting inertia ratio is $J_L/J_M = 1.75$, preventing control instability.

The support structure consists of modular aluminium profiles integrated into a base frame that interfaces directly with the wind-tunnel metal framework, see Figure 7 (right). Reinforced plates and cross-bracing ensure that structural vibrations remain low.

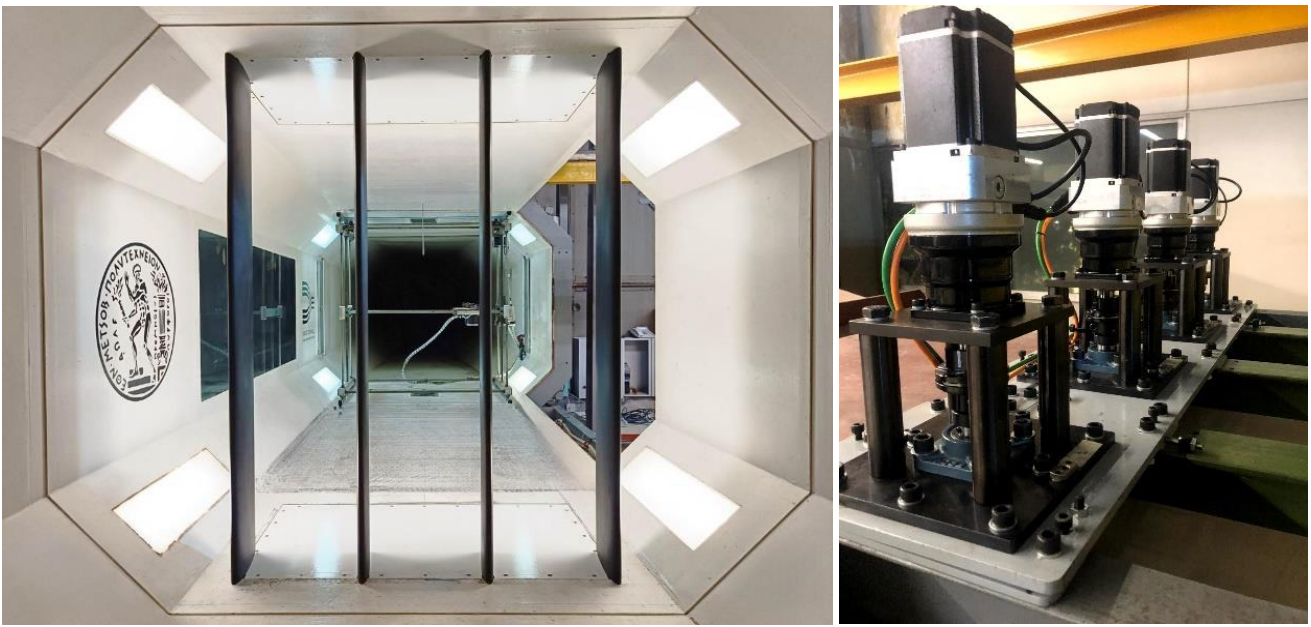

*Figure 7. (left) Downstream view inside the test section with Gust Generator the vanes installed. (Right) Detail of the four servo motors installed at the top, outside the test section.*

### 3.1.4 Measurements

Two component velocity measurements were performed in the wake of the GG with a crossflow hot wire probe (TSI 1243). The probe was calibrated using a TSI Model 1127 Air Velocity Calibrator connected to the labs pressurized air circuit. Initially 17 velocity points were recorded with the probe aligned to the flow followed by yaw calibration. The latter was performed at five reference velocities and for each flow velocity the probe sensors were sampled across 11 yaw positions, (5 × 11 data samples) spanning a yaw angle range of [-30°, 30°] in 6° increments. The TSI IFA 300 Constant Temperature Anemometer system was used for the calibration and measurements. The final Mean Absolute Error obtained for the velocity and flow angle were $\varepsilon_U = 1.08\%$, $\varepsilon_\theta = 7.11\%$, respectively.

The measurement campaign concerned measurements at two points, A and B, located 10 vane chords ($c = 0.2\ m$) downstream of the GG vanes, see Figure 8. Points $A\ (10c, -0.18c)$ and $B\ (10c, -0.85c)$, were approximately at the centre of the wind tunnel and in the wake of the third vane, respectively. Both points were at the mid-span level of the vanes, which corresponds to the test section's mid-height. A 4th-order Butterworth low-pass filter with a 100 Hz cutoff frequency was used to remove high-frequency noise while preserving low-frequency signal components. The filter was applied bi-directionally to eliminate phase shift. To mitigate structural disturbances in the gust measurements, a zero-phase, narrowband, high-quality-factor infinite impulse response notch filter was applied at the eigenfrequency of the probe support structure and its second and third harmonics. Subsequently, harmonic-regression subtraction was employed to further attenuate residual oscillatory components.

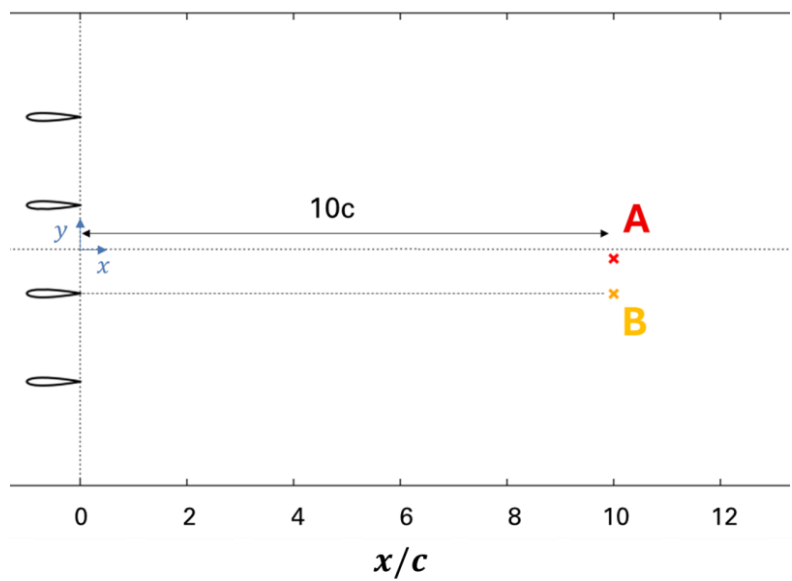

Figure 8. Schematic of the Gust Generator Characterisation set up and measurement point

### 3.1.5 Examined Gusts

The present study considers '1-cos' gusts. The specific profile has been systematically shown in the literature to produce undesirable negative peaks in gust velocity and gust angle due to the formation of starting and stopping vortices, e.g. [38].

The gust profile is generated by two methods. First by having the vanes follow an identical motion as the nominal gust. The discrete motion is defined as follows:

$$\theta(t) = \begin{cases} 0 & , for\ 0 \leq t \leq t_{01} \\ \dfrac{A}{2} \cdot \left(1 - \cos\left(2\pi f_{gust}(t - t_{01})\right)\right) & , for\ t_{01} < t \leq t_{01} + \dfrac{1}{f_{gust}} \\ 0 & , for\ t > t_{01} + \dfrac{1}{f_{gust}} \end{cases} \quad (4)$$

where $A$ is the vane motion amplitude, $f_{gust}$ the vane motion frequency and $t_{01}$ the time instance when the motion starts.

Second, a modified vane motion profile is considered, following the work of [34]. The modified motion equation is defined across three distinct phases, as shown in equation (5):

$$\theta(t) = \begin{cases} e^{bt} - 1 & , for\ 0 \leq t \leq t_1 \quad \text{(Exponential rise)} \\ \dfrac{A}{2} \cdot \left(1 - \cos(2\pi f(t - t_{01}))\right) & , for\ t_1 < t \leq t_2 \quad \text{(Harmonic motion)} \\ e^{c(t - t_{02})} & , for\ t > t_2 \quad \text{(Exponential decay)} \end{cases} \quad (5)$$

where $b, c$ are user defined parameters. This is also true for the time instances $t_{01}, t_{02}, t_1, t_2$.

Following a comprehensive numerical investigation [39], the values given in Table 3 were selected for the modified vane motion. Compared to [34], $\theta(t_2)$ is increased from $0.5A$ to $0.7A$ and $t_3$ is increased from $t_{01} + 5/f_{gust}$ to $t_{01} + 7/f_{gust}$. Numerical predictions from [39] suggest this leads to a 55% reduction in the negative peak factor following the peak gust angle, compared to [34].

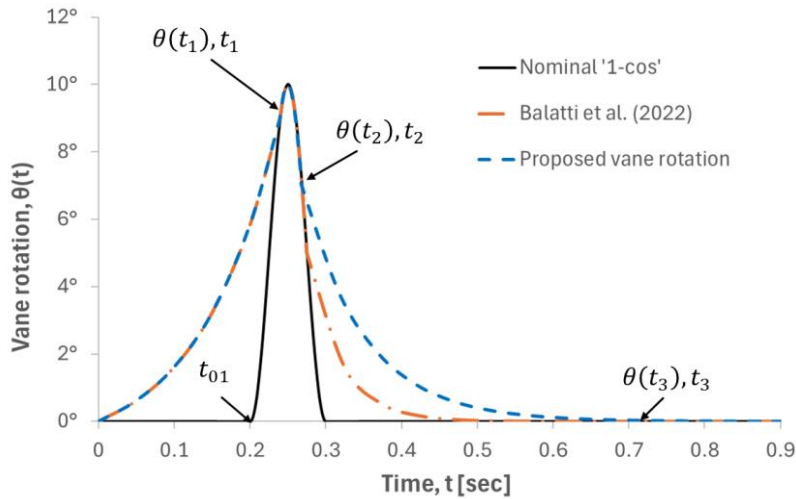

Figure 9. Parameterized vane motion to improve the generated gust with respect to the nominal '1-cos' motion and the literature [34].

Table 3. Selected parameters for the parameterized vane motion, see equation 5.

| Parameter | $A = 10°$ | $A = 20°$ |
|---|---|---|
| $t_{01}$ [s] | 0.20 | 0.30 |
| $t_3$ [s] | $t_{01} + 7/f_{gust}$ | $t_{01} + 7/f_{gust}$ |
| $\theta°(t_1)$ | $0.9A$ | $0.9A$ |
| $\theta°(t_2)$ | $0.7A$ | $0.7A$ |
| $\theta°(t_3)$ | $0.003°$ | $0.003°$ |

In total, sixteen test cases were considered, see Table 4, including two vane motion profiles ('1–cos' and modified). For each motion profile, experiments were performed at freestream velocities of 10 and 20 m/s. At each velocity, vane oscillation amplitudes of 10° and 20° were tested in combination with forcing frequencies of 10 Hz and 20 Hz, yielding eight cases per motion profile. In the interest of brevity and without any loss in the generality of the conclusions of this study, only selected results are shown in the subsequent graphs. The reduced frequency defined as $k = \frac{\pi f_{gust} c}{U_\infty}$ is also given for each case. The reduced frequency for all cases is $k > 0.2$ setting the aerodynamic problem in the highly unsteady regime.

*Table 4. Test cases for Gust Generator Characterisation*

| Test Case | Vane motion profile | $U_\infty$ [m/s] | A | $f_{gust}$ [Hz] | $k$ |
|---|---|---|---|---|---|
| 1 | '1-cos' | 10 | 10° | 10 | 0.63 |
| 2 | '1-cos' | 10 | 20° | 10 | 0.63 |
| 3 | '1-cos' | 10 | 10° | 20 | 1.26 |
| 4 | '1-cos' | 10 | 20° | 20 | 1.26 |
| 5 | '1-cos' | 20 | 10° | 10 | 0.31 |
| 6 | '1-cos' | 20 | 20° | 10 | 0.31 |
| 7 | '1-cos' | 20 | 10° | 20 | 0.63 |
| 8 | '1-cos' | 20 | 20° | 20 | 0.63 |
| 9 | modified | 10 | 10° | 10 | 0.63 |
| 10 | modified | 10 | 20° | 10 | 0.63 |
| 11 | modified | 10 | 10° | 20 | 1.26 |
| 12 | modified | 10 | 20° | 20 | 1.26 |
| 13 | modified | 20 | 10° | 10 | 0.31 |
| 14 | modified | 20 | 20° | 10 | 0.31 |
| 15 | modified | 20 | 10° | 20 | 0.63 |
| 16 | modified | 20 | 20° | 20 | 0.63 |

## 3.2  Computational approach

### 3.2.1  Numerical Framework

All simulations were carried out with MaPFlow [40], the in-house CFD framework developed at NTUA. For the present work, its Eulerian module was employed. This is a compressible, cell-centred solver capable of operating on both structured and unstructured meshes. The code provides a range of turbulence closures, including the Spalart–Allmaras (SA) one-equation model [41] which is mainly employed in this study. MaPFlow also incorporates and a Delayed Detached-Eddy Simulation (DDES) formulation based on the SA model as proposed in [42].

In this work, the wind tunnel walls are also included in the simulations. Consequently, to accommodate the relative movement of the gust generators vanes with respect to the tunnel work a deforming mesh methodology is incorporated. Mesh deformation is based on a Radial Basis Function (RBF) approach that results in smoother mesh deformation. For more information the reader can refer to [43,44]. At each time step, the computational domain is deformed to accommodate the movement of the solid surfaces. The new position of the mesh's internal nodes is calculated using the interpolation function of Equation (6):

$$s(\vec{x}) = \sum_{i=1}^{N} \alpha_i \varphi(\|\vec{x} - \vec{x}_{si}\|) + p(\vec{x}) \tag{6}$$

where $s(\vec{x})$ is the interpolation function of an internal node $\vec{x}$, $\alpha_i$ the interpolation weights and $\varphi$ the RBF function, see Equation 7. The latter depends on the normed distance between the internal node $\vec{x}$ and the wall nodes $\vec{x}_{si}$. The RBF function chosen for the current study was the Wendland C2 [45]. Finally, $p(\vec{x})$ is a first order polynomial given by Equation 8.

$$\varphi(\|\vec{x}\|) = \begin{cases} (1 - \|\vec{x}\|)^4 (4\|\vec{x}\| + 1) & \|\vec{x}\| < 1 \\ 0 & \|\vec{x}\| \geq 1 \end{cases} \tag{7}$$

$$\begin{aligned} p^x(\vec{x}) &= \gamma_o^x + \gamma_x^x x + \gamma_y^x y + \gamma_z^x z \\ p^y(\vec{x}) &= \gamma_o^y + \gamma_x^y x + \gamma_y^y y + \gamma_z^y z \\ p^z(\vec{x}) &= \gamma_o^z + \gamma_x^z x + \gamma_y^z y + \gamma_z^z z \end{aligned} \tag{8}$$

The Support Radius for the Deformation method was chosen to be 350 mm (1.75c) and to ensure that no conflicts between the gust generators exist and conflicts with the domain boundaries, the deformation was exponentially decayed after the first 150 mm (0.75c).

### 3.2.2 Mesh and Timestep Independence Study

The basic computational setup can be found in Figure 10. The tunnel walls are modelled as slip walls, while the four GG vanes are treated as no-slip walls. In the vicinity of the vanes as well as in the wake region there is a refinement zone to properly capture the generated vortices. For all the present analysis the SA one equation turbulence model is adopted.

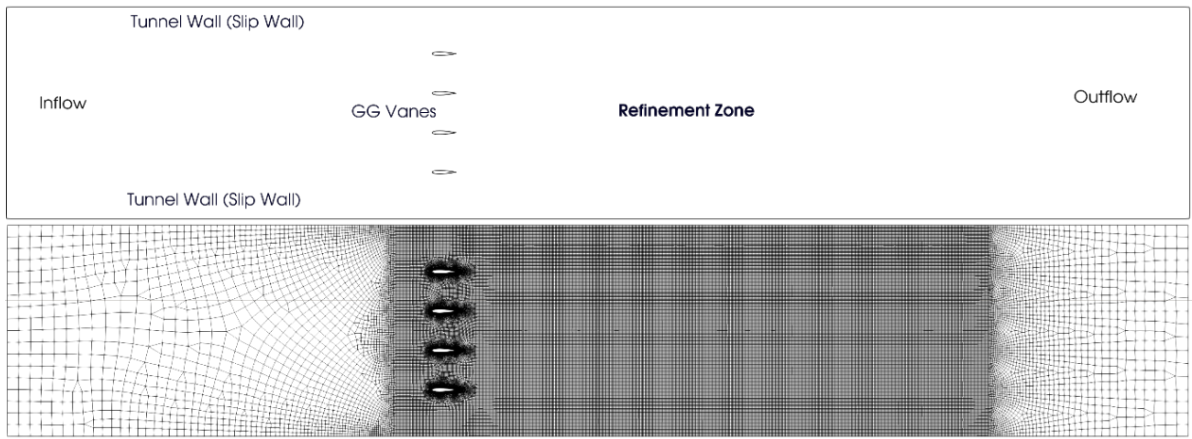

*Figure 10. Basic Computational Setup*

For the grid and time-step sensitivity analysis, three mesh resolutions, coarse, medium, and fine, were examined in conjunction with three time-step sizes corresponding to 500, 1000, and 4000 time steps per period. For the mesh details, see Table 5.

To quantify the computational differences arising from variations in numerical parameters, velocity profiles are examined for a harmonic oscillation of the GG vanes at 10 Hz with an amplitude of 10° and a free-stream velocity $U_\infty = 20 m/s$. A monitor point is located 1200 mm (six gust-generator chord lengths) downstream of the vanes' trailing edge, at the centre of the wind tunnel width. Flow velocities recorded at this location are used for comparison across the different numerical configurations.

Table 5. Mesh details for the grid independence study.

| Grid | Number of Cells | Y+ | Refinement Zone Cell Size |
|---|---|---|---|
| Coarse | 184500 | <1 | 20mm |
| Medium | 379054 | <1 | 10mm |
| Fine | 518709 | <1 | 5mm |

Regarding mesh resolution, it is evident from Figure 11, left, that the medium mesh can adequately capture the velocity fluctuations downstream of the vanes. Regarding the time step resolution, it is clear that 500 steps per period cannot sufficiently resolve the flow. For the higher time resolution, only minor differences can be identified, consequently a time step resolution of 1000 steps per period is selected. To that end, for all the following numerical studies the medium mesh will be employed with a timestep corresponding to 1000 steps per period.

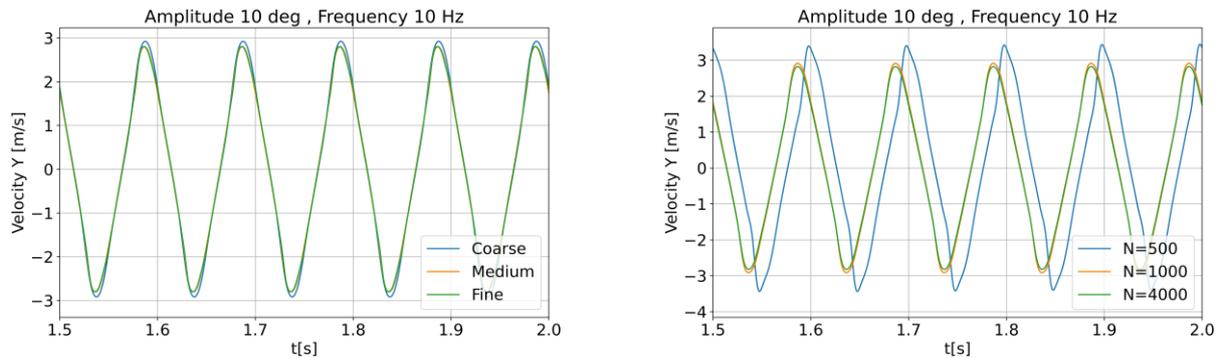

Figure 11 Grid (Left) and Time-step (Right) Independence study

### 3.2.3 Modelling Strategy

Following the grid independence study, a numerical investigation is conducted to quantify how numerical modelling choices affect the simulation results. Specifically, the two-dimensional assumption is questioned and the option of a higher fidelity turbulence model (DDES) is explored.

Initially, regarding the 2D assumption, a simple 3D domain was constructed by extruding the existing 2D mesh to an Aspect Ratio (AR) of one. The setup is similar to the one presented in Figure 10, while all the side boundaries are treated as inviscid walls (slip condition). Using the above setup both 3D

URANS and DDES simulations are conducted to examine whether three dimensional effects are present when the GG vanes are operating. For all the 3D simulations the SA model is exploited using either the RANS variant or the higher fidelity DDES one [42].

To compare the differences between the various setups the same monitor point, six chords downstream of the GG vanes, is used. Velocity fluctuations are recorded throughout the simulations and are compared for the various models. The metric delta, $d_{method}$, is introduced to quantify the differences between methodologies, as defined in Equation 9

$$d_{method} = \frac{u_{method} - u_{3D_{RANS}}}{U_\infty} \tag{9}$$

where the index method can be either '2D,URANS' or 'DDES', and $u$ can be any velocity component. The velocity variations and their differences can be found in Figure 12 and Figure 13, respectively.

Results show that the solutions are largely insensitive to the modelling strategy employed, with only minor differences observed. The 3D methodologies exhibit only small deviations, up to $0.005 U_\infty$, from the two-dimensional, lower fidelity approach. Furthermore, practically no discrepancies are identified between the 3D URANS and DDES simulations.

For the low-Mach-number, confined oscillating-vane configurations examined here, the 3D URANS simulations are shown to capture the essential gust-generation and propagation physics with sufficient fidelity. Consequently, the 3D URANS approach was selected over the 2D RANS formulation in order to account for potential three-dimensional flow effects while maintaining computational efficiency. Extensions to configurations involving strong end-wall effects, spanwise-nonuniform actuation, or substantially higher Reynolds numbers would require additional assessment but do not affect the conclusions drawn for the present class of flows.

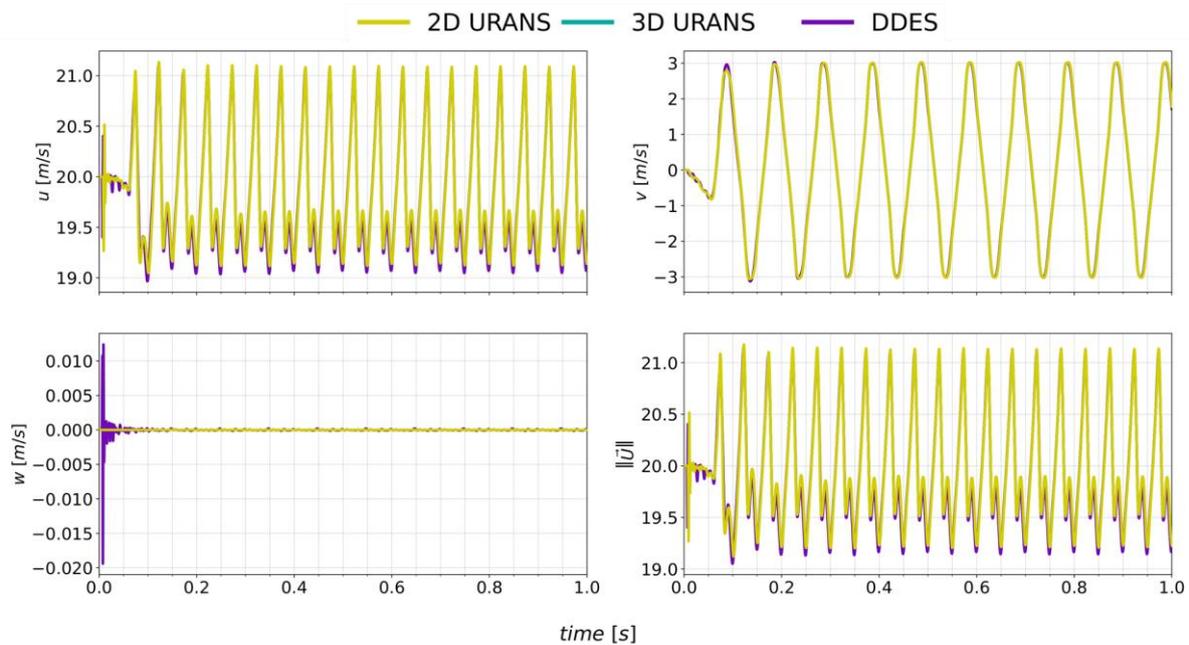

*Figure 12. Velocity components over simulation time. The velocity components and magnitude are plotted for the cases of 2D URANS, 3D URANS and DDES simulations.*

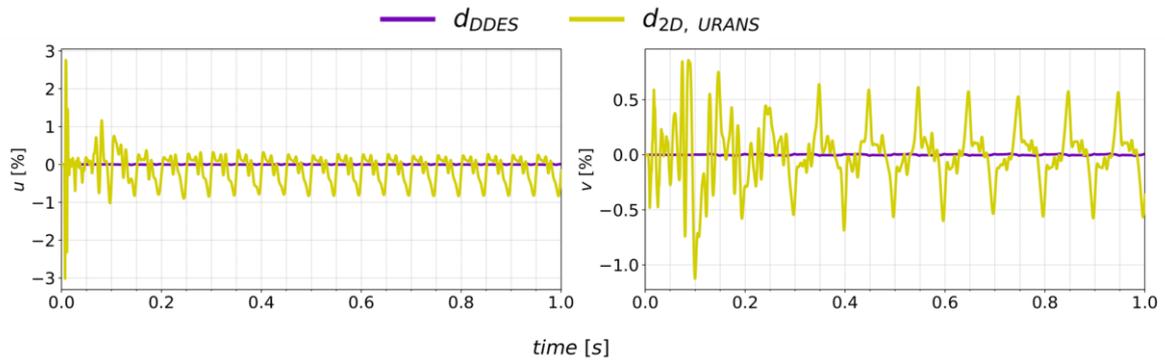

*Figure 13. Velocity's normalized differences. Having the 3D RANS as a base case, the u, v components of the other cases' velocity are subtracted to the base case's respective component. Then, the velocity differences are normalized with the Freestream Velocity.*

# 4 Results and Discussion

## 4.1 Experiments

The performance of the new gust generator was evaluated experimentally under a range of operating conditions. Figures from 14 to 17 show the effect of free stream velocity, vane motion amplitude, vane motion frequency and motion profile. In all figures, t=0 corresponds to the start of the motion for the '1-cos' protocol. The modified motion starts 0.2 s earlier, see Figure 9.

Starting with a nominal '1-cos' input, Figure 14 shows the effect of increasing the free-stream velocity. As the wind speed is raised from 10 m/s to 20 m/s, the gust arrives earlier at the sensor location, as expected. Given the 2 m distance between the GG vane trailing edge and the sensor, the convection time decreases from approximately 0.2 s at 10 m/s to 0.1 s at 20 m/s. The disturbance in the flow appears to be convected downstream at the free stream velocity. GR remains largely unchanged, or, in the higher-amplitude case, is slightly reduced at higher velocities. Additionally, higher free-stream velocity produces more pronounced negative peaks in the measured gust profile.

The effect of increasing the vane motion amplitude for the same '1-cos' gust input is shown in Figure 15. Increasing the amplitude, $A$, from 10° to 20° approximately doubles the gust amplitude at 10 m/s and produces even higher increases at 20 m/s. Further, increasing $A$ from 10° to 20° at both tested velocities results in more pronounced negative peaks, and stronger secondary effects. At the lower amplitude, the generated gust exhibits a hat-like shape potentially affected by the secondary velocity variations.

Figure 16 presents the effect of increasing vane motion frequency $f_{gust}$ from 10 Hz to 20 Hz. The increase in $f_{gust}$ reduces the overall measured gust duration, as expected because of the shorter actuation cycle. GR also increases with frequency, with the enhancement being more pronounced at the higher free-stream velocity. The magnitude of the negative peaks remains largely unchanged across the two frequencies, while the associated secondary effects become less prominent at the higher actuation frequency.

The effect of the modified vane motion is shown in Figure 17. The gust generated by the proposed motion protocol exhibits reduced negative excursions and a smaller peak gust amplitude compared to the nominal case. This change in motion leads to a significant reduction in NPF, while the GR is also affected, as expected due to the smoother actuation. Secondary effects are still observable for both motion protocols.

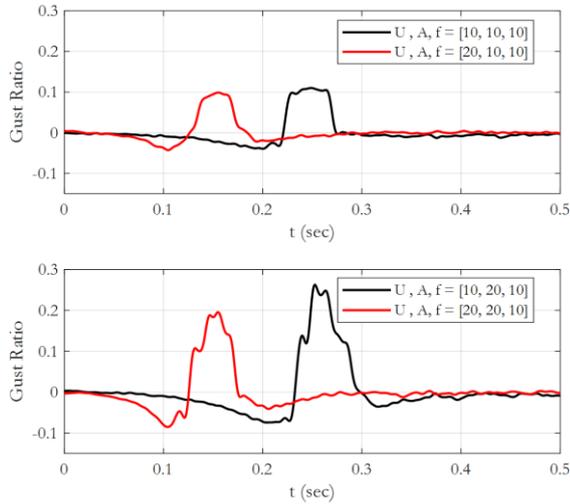

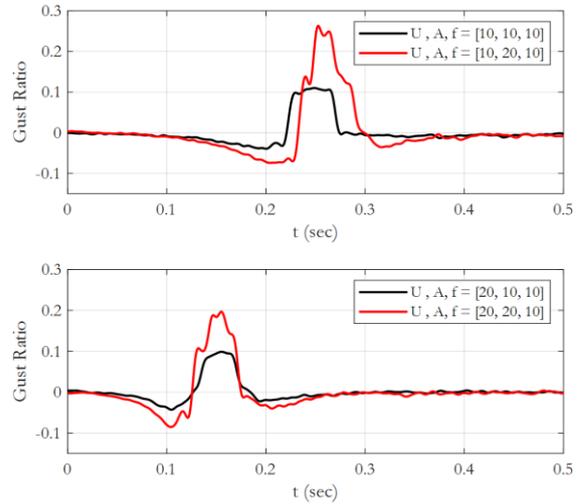

Figure 14. Effect of increasing Free Stream Velocity for a nominal '1-cos' vane motion with $f_{gust} = 10\ Hz$ and two amplitudes $A_{vane} = 10°$ (top) and $A_{vane} = 20°$ (bottom). Comparison between $U_\infty = 10\ m/s$ and $U_\infty = 20\ m/s$.

Figure 15. Effect of increasing motion Amplitude for a nominal '1-cos' vane motion with $f_{gust} = 10\ Hz$ and two free stream velocities, $U_\infty = 10\ m/s$ (top) and $U_\infty = 20\ m/s$ (bottom). Comparison between $A_{vane} = 10°$ and $A_{vane} = 20°$.

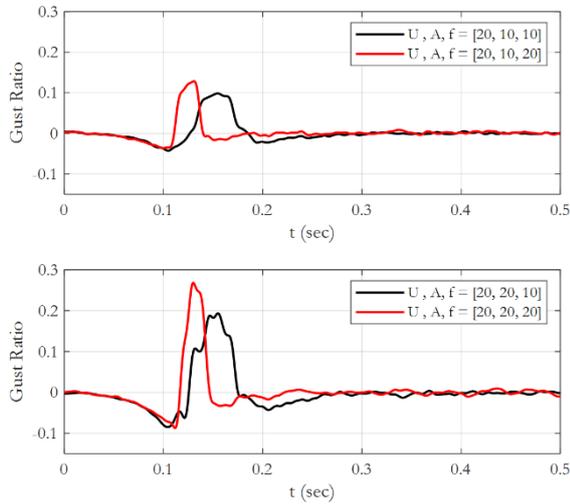

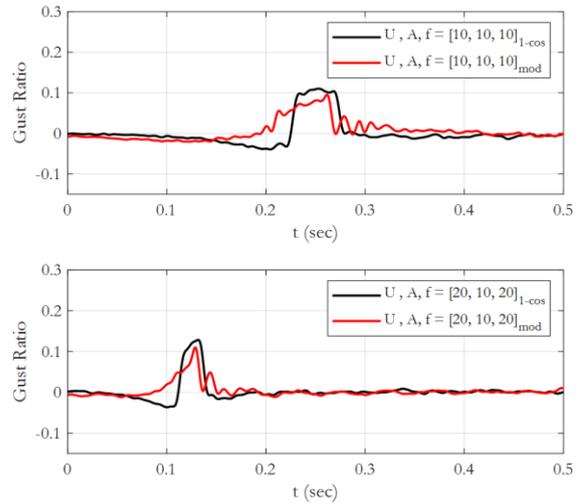

Figure 16. Effect of increasing Vane frequency for a nominal '1-cos' vane motion at a free stream velocity of $U_\infty = 20\ m/s$ and for two Amplitudes, $A_{vane} = 10°$ (top) and $A_{vane} = 20°$ (bottom). Comparison between $f_{gust} = 10\ Hz$ and $f_{gust} = 20\ Hz$.

Figure 17. Effect of Vane motion protocol for a vane amplitude $A_{vane} = 10°$ at two free stream velocities, $U_\infty = 10\ m/s$ (top) and $U_\infty = 20\ m/s$ (bottom) and two frequencies $f_{gust} = 10\ Hz$ (top) and $f_{gust} = 20\ Hz$ (bottom).

The achieved NPFs for the nominal '1-cos' and the proposed motion are given in Table 6 for two operating conditions, $[U, A, f] = [10,10,10]$ and $[20, 10, 20]$. For both conditions, the proposed motion yields lower NPF and GR values than the original 1-cos profile, indicating a strong reduction in negative velocity peaks and a milder reduction in gust intensity. The reductions are more pronounced at the higher amplitude and frequency condition, suggesting improved performance of the proposed motion under more aggressive forcing.

Table 6. Negative Peak factor values achieved for the nominal '1-cos' and the proposed motion profile

| Conditions | [U, A, f] = [10, 10, 10] | | [U, A, f] = [20, 10, 20] | |
|---|---|---|---|---|
| Vane motion profile | Negative peak factor (NPF) | Gust Ratio (GR) | Negative peak factor (NPF) | Gust Ratio (GR) |
| Original '1-cos' | 0.36 | 0.11 | 0.27 | 0.13 |
| Proposed motion | 0.23 | 0.09 | 0.13 | 0.10 |
| Change | -37% | -17% | -54% | -20% |

It is evident from the presented results and the large number of parameters involved that a non-dimensional analysis would provide greater insight. Figure 18 shows the GR variation with reduced frequency, $k = \frac{\pi f c}{U_\infty}$, for all cases in Table 4 and for both measurement points. The results confirm that the proposed motion does provide a robust way to limit NPF with minimal effects on peak Gust Ratio. Also increasing $k$ beyond $k = 0.6$ has little effect on the achieved GR or NPF.

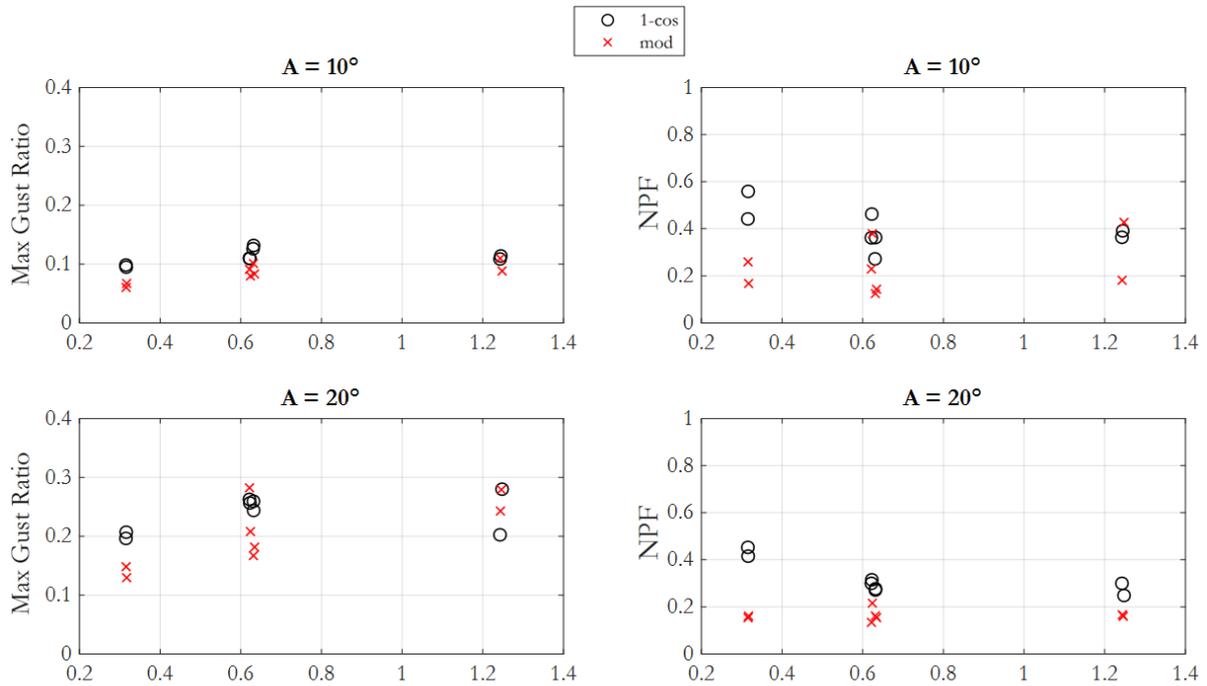

Figure 18. Maximum Gust Ratio and Negative Peak Factor for all cases in Table 4, for both measurement points.

## 4.2 Numerical investigation

### 4.2.1 Validation against measurements

The numerical predictions are compared against point measurements from the wind tunnel tests. The comparisons are presented in Figure 19 and Figure 20 for vane amplitudes of 10° and 20°, respectively, as a validation exercise. Cases concern both 10 and 20 m/s free stream velocities. The comparison shows excellent agreement between simulations and measurements in terms of positive and negative gust peaks and timings. As expected, for the higher free stream velocity cases the gust is convected quicker downstream and this is accurately captured by the simulations. It is also noted that in both experiments and simulations secondary variations are observed, see vectors in Figure 19 and Figure 20. These secondary variations, discussed further in the following section, are more pronounced in the high amplitude case and are well captured in the simulations, confirming the flow is modelled accurately.

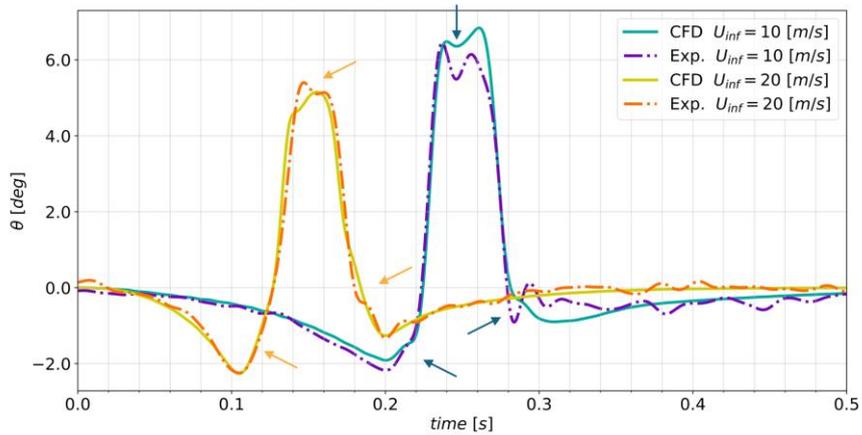

Figure 19. Gust angle at point A for '1-cos' vane motion, vane amplitude $A = 10°$ and vane frequency $f_{gust} = 10\ Hz$. Comparison between measurements and 3D RANS simulations.

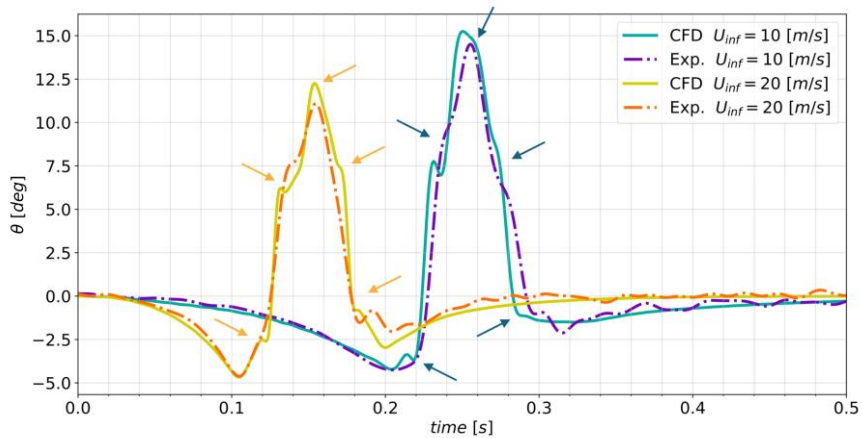

Figure 20. Gust angle at point A for '1-cos' vane motion, vane amplitude $A = 20°$ and vane frequency $f_{gust} = 10\ Hz$. Comparison between measurements and 3D RANS simulations.

### 4.2.2 Numerical investigation of vane motion effect

In this section, the original '1-cos' motion and the modified vane motion are compared in terms of gust propagation using the extracted flow field from the numerical simulations. The freestream velocity is set at 20 m/s, while two gust generation configurations are examined: a 'mild' case with a frequency of 10 Hz and amplitude of 10°, and an 'intense' case with a frequency of 20 Hz and amplitude of 20°.

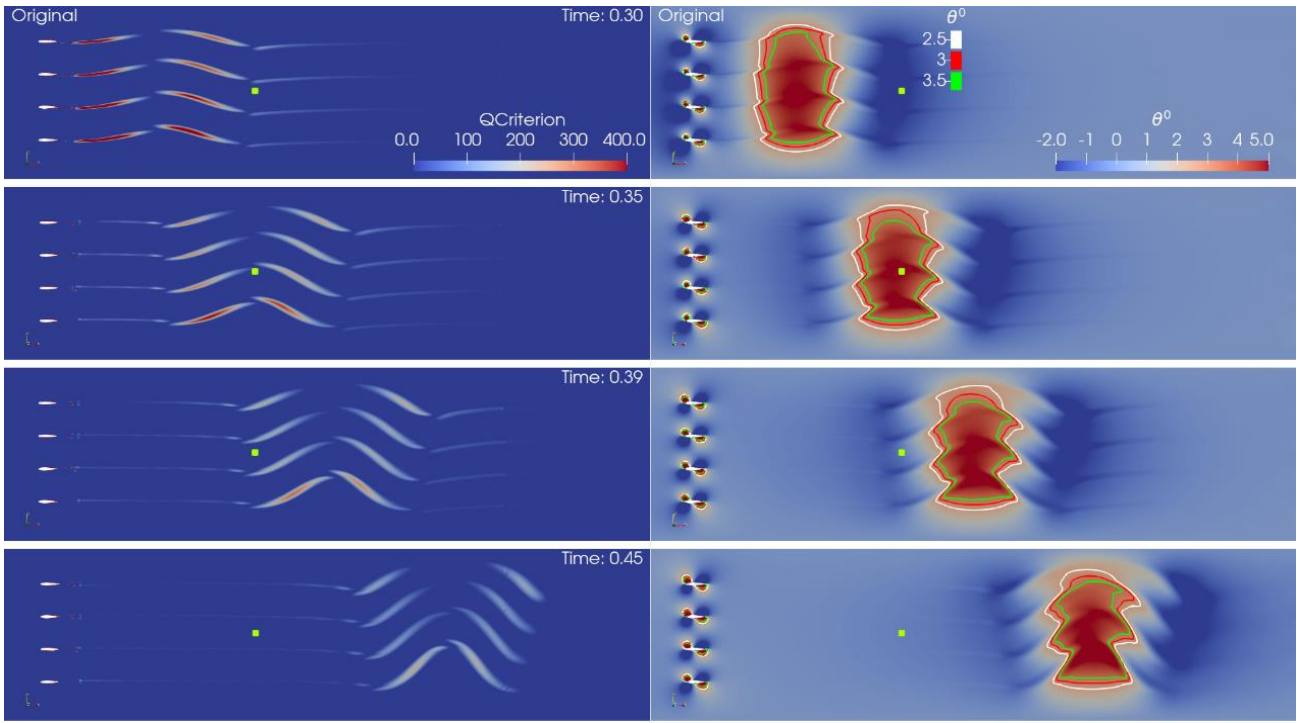

*Figure 21. Gust propagation for the original motion. (Left) Q criterion contours at different time stamps; (Right) flow angle contours with isolines at different time stamps. Interaction between the discrete vortices produces secondary variations, which are evident as asymmetries in the flow angle ($U_\infty = 20\frac{m}{s}, f_{gust} = 10\ Hz, A = 10°$). The green square displays the location of monitor point A. It is noted that the whole computational domain is shown with the top and bottom walls included.*

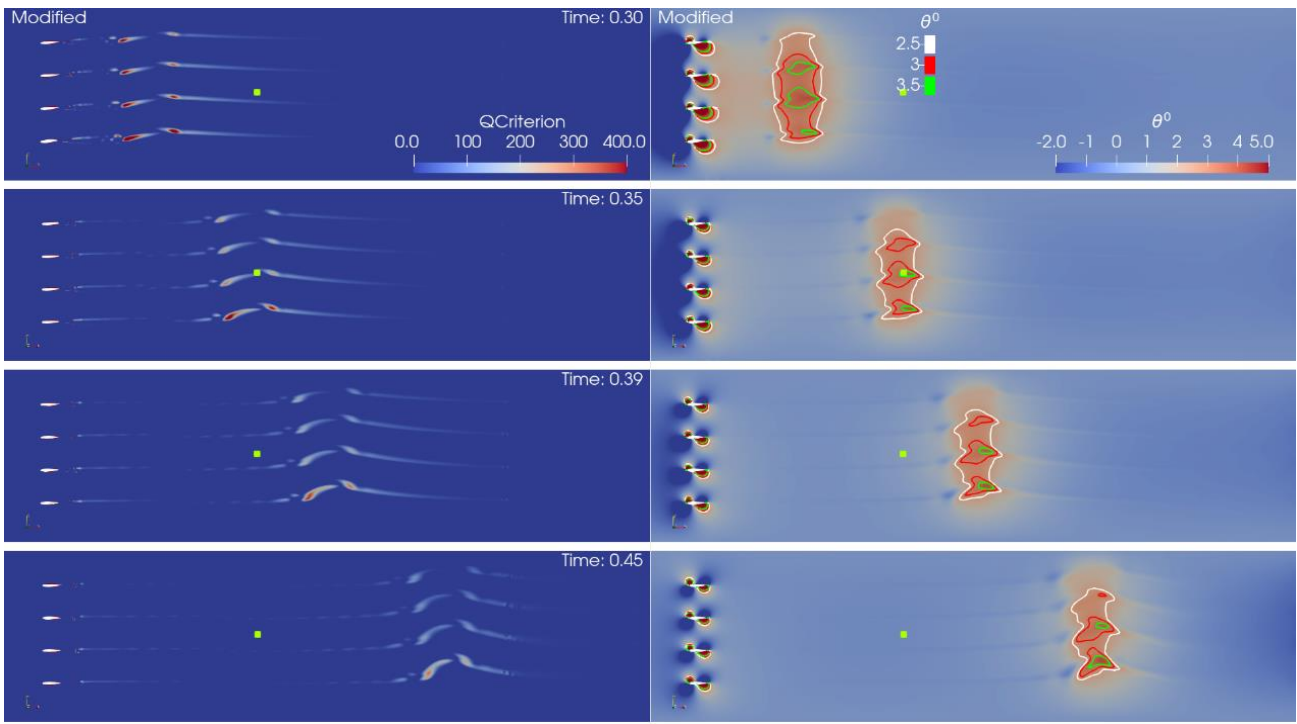

*Figure 22. Gust propagation for the modified motion. (Left) Q criterion contours at different time stamps; (Right) flow angle contours with isolines at different time stamps. Interaction between the discrete vortices produces secondary variations, which are evident as asymmetries in the flow angle ($U_\infty = 20\frac{m}{s}, f_{gust} = 10\ Hz, A = 10°$). The green square displays the location of monitor point A. It is noted that the whole computational domain is shown with the top and bottom walls included.*

Figure 21 illustrates the q-criterion and flow angle ($\vartheta$ in degrees) contours for the original '1-cos' motion. The Q criterion contours, Figure 21-left, reveal that each vane generates two strong elongated discrete vortices, which essentially generate the gust peak. The distance between the elongated vortices at the top of the figure increases as they are propagated downstream, and each vortex pair appears to engulf the vortex pair below. The two main vortices are preceded by two weaker vortices, the start and stop vortices, which are responsible for the negative peaks ahead and after the measured gust. The effect of the top wall is evident as the vortices propagate downstream, with the variation across the wind tunnel y axis. Examination of the isolines for selected θ values, Figure 21-right, reveals distinct asymmetries in the flow field. These asymmetries manifest as secondary variations in the flow angle and are attributed to the interactions between the wakes generated by adjacent vanes.

Figure 22 presents the same data for the modified motion. The discrete vortices are now rounder, smaller and weaker compared to the '1-cos' case. This explains the weaker gust peaks generated by the modified motion protocol. The 'start' and 'stop' vortices are also milder, due to the reduced acceleration and deceleration of the vanes, which leads to the reduced negative peaks. The top wall effect is also clearly visible in this case. No engulfing interaction is visible, unlike the '1-cos' case.

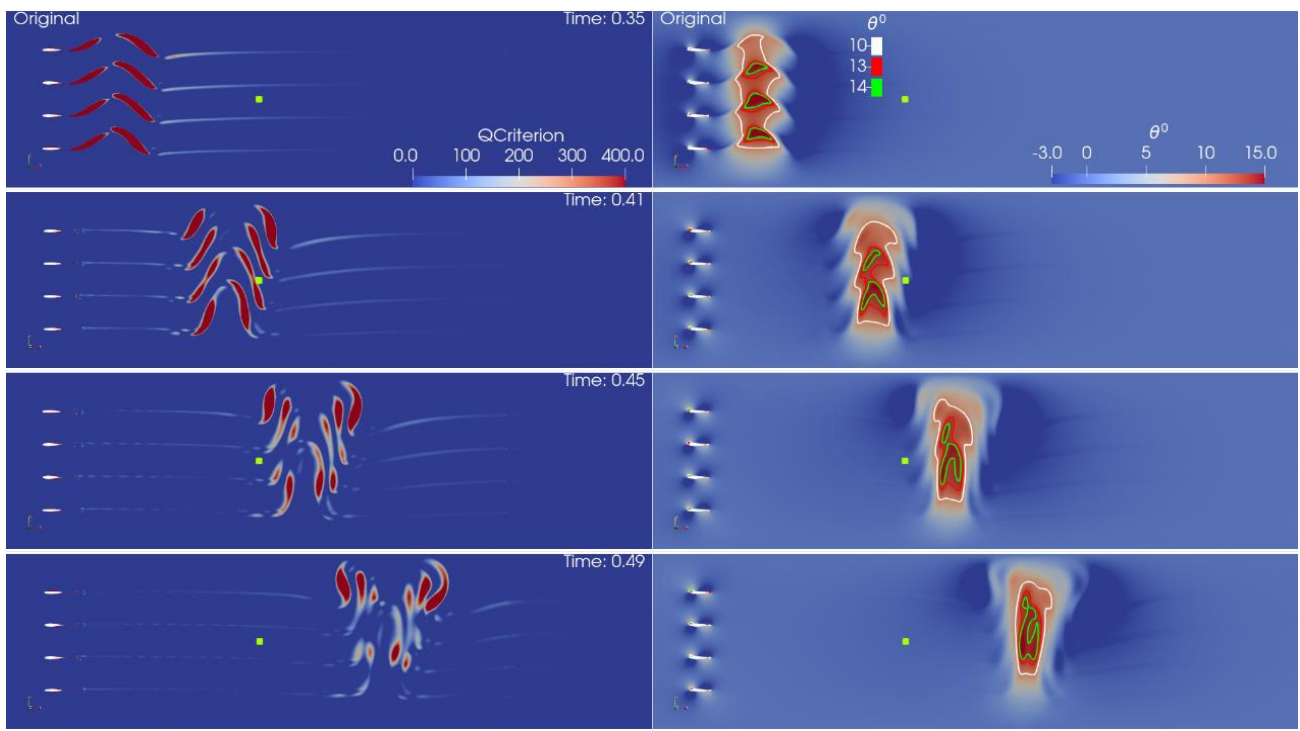

*Figure 23. Gust propagation for the original motion. (Left) Q criterion contours at different time stamps; (Right) flow angle contours with isolines at different time stamps. Interaction between the discrete vortices produces secondary variations, which are evident as asymmetries in the flow angle ($U_\infty = 20\frac{m}{s}, f_{gust} = 20\ Hz, A = 20°$). The green square displays the location of monitor point A. It is noted that the whole computational domain is shown with the top and bottom walls included.*

The comparison for the stronger excitation case (frequency of 20 Hz, Amplitude 20°) is shown in Figure 23 and in Figure 24, which show Q criterion and θ isolines for the original and modified motion, respectively. Consistent with the mild case, the original motion generates significantly stronger, more elongated gusts, a characteristic that is even more pronounced at this higher frequency and amplitude. The vortex interactions between vanes are also intensified and more complex, with the elongated vortices being split for the lower vanes wakes, see Figure 23-left for $t \geq 0.41$. This leads to more pronounced secondary variations in the flow field, see θ isolines in Figure 23-right. Flow angle variation normal to the streamwise direction is still significant.

For the modified motion, rounder weaker vortices are observed, Figure 24-left. The starting and stopping vortices are now clearly visible. Interestingly, the modified motion protocol exhibits a varying propagation speed across the tunnel width, evident in both the vorticity contours and the flow angle distribution. This gust "inclination" is more apparent for $t \geq 0.41$ and is conceivably a combined effect of top wall confinement and wake vortex interaction, see Figure 24-left. The asymmetry of the generated vortex is amplified as the gust propagates downstream.

Regarding the secondary variations, these are of course substantial in the 'stronger' case. Indeed, as Figure 23 suggests, wakes generated by adjacent vanes interact with each other, leading to small but distinct changes on the flow angle distribution.

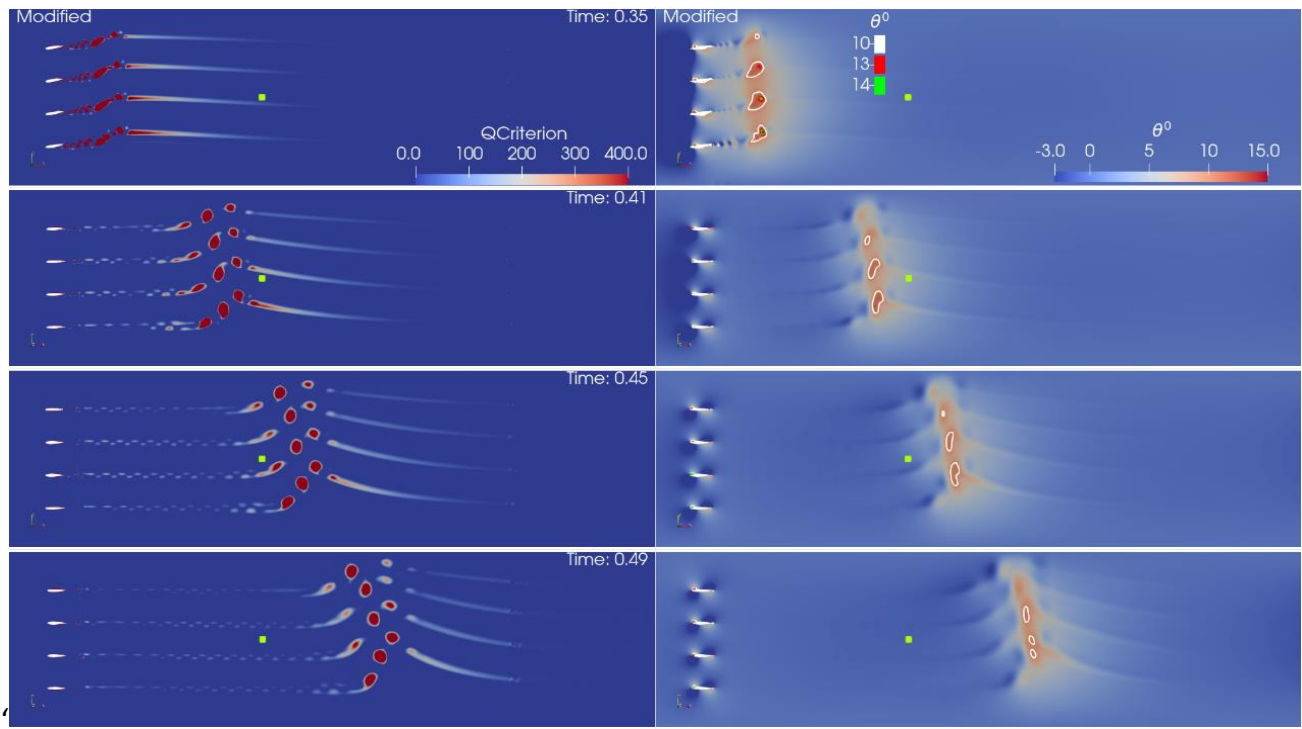

*Figure 24. Gust propagation for the modified motion. (Left) Q criterion contours at different time stamps; (Right) flow angle contours with isolines at different time stamps. Interaction between the discrete vortices produces secondary variations, which are evident as asymmetries in the flow angle ($U_\infty = 20 \frac{m}{s}, f_{gust} = 20\ Hz, A = 20°$). The green square displays the location of monitor point A. It is noted that the whole computational domain is shown with the top and bottom walls included.*

## 5 Conclusions

In this work, a four-vane oscillating gust generator has been designed, manufactured, and characterised for installation in the NTUA low-speed wind tunnel. The system achieves gust ratios relevant to aircraft, UAV, and wind-turbine applications while maintaining mechanical robustness and precise motion control. A combined experimental and numerical approach demonstrated that a three-dimensional URANS modelling strategy provides sufficient fidelity for capturing the essential gust dynamics and accurately reproduces measured velocity and flow-angle histories.

A modified vane motion protocol was introduced to address the pronounced negative velocity peaks inherent to nominal '1-cos' gusts. The reduction in negative peak factor is achieved through weakened start–stop vortices associated with reduced vane acceleration, implying an intrinsic trade-off between gust sharpness and peak amplitude. Both measurements and simulations show that this approach significantly reduces the negative peak factor, with only a moderate reduction in peak gust ratio, and that the improvement becomes more pronounced under stronger excitation conditions.

Similarly, both experiments and simulations reveal secondary variations in flow angle. CFD flow-field analysis showed that these originate from nonlinear interactions between vortices shed by adjacent vanes and from confinement effects imposed by the wind-tunnel walls. At higher reduced frequencies, the dominance of coherent vortex interactions limits further gains in gust strength and amplifies secondary flow-angle distortions.

Overall, the present gust generator provides a high-quality experimental platform for systematic investigations of gust–airfoil and gust–structure interaction in highly unsteady regimes. The insights gained into gust formation, propagation, and secondary flow effects also offer guidance for the design of future gust generators and for the interpretation of unsteady aerodynamic experiments.

**Acknowledgements:**

The authors extend their gratitude to Sotiris Mavrakis, technical personnel at NTUA wind Tunnel, for his support during the experiments for this study.

This project is carried out within the framework of the National Recovery and Resilience Plan Greece 2.0, funded by the European Union – NextGenerationEU (H.F.R.I. Project Number: 016749).

The meshes were generated by using the ANSA BETA CAE preprocessor. All the computations were performed using the VEGA HPC under the EuroHPC project **HPCAGREE** with id EHPC-REG-2024R01-058.